\documentclass[10pt,twocolumn,showpacs,superscriptaddress,prl]{revtex4}

\usepackage[tbtags]{amsmath}
\usepackage{amsthm}
\usepackage{amssymb}
\usepackage{amsfonts}
\usepackage{bbm,bm,mathrsfs}
\usepackage{graphicx}
\usepackage{xcolor}
\usepackage{hyperref}
\hypersetup{colorlinks=true,
            breaklinks=true,
            pdfstartview=FitH,
            linkcolor=blue,
            anchorcolor=red,
            citecolor=blue}

\allowdisplaybreaks

\DeclareMathOperator{\Tr}{Tr}%
\DeclareMathOperator{\diag}{diag}%

\newcommand{\bra}[1]{\langle #1\vert}
\newcommand{\ket}[1]{\vert #1\rangle}

\theoremstyle{plain}
\newtheorem{lemma}{Lemma}
\begin{document}
\title{Quantum Discord: A Dynamic Approach in Geometric Picture}

\author{Mingjun Shi}%
\email{shmj@ustc.edu.cn}%
\affiliation{Department of Modern Physics, University of Science and Technology of China, Hefei,
Anhui 230026, People's Republic of China}

\author{Fengjian Jiang}
\affiliation{Department of Modern Physics, University of Science and Technology of China, Hefei,
Anhui 230026, People's Republic of China}
\affiliation{Huangshan University, Huangshan, Anhui 245021, People's Republic of China}

\author{Jiangfeng Du}
%\email{djf@ustc.edu.cn}%
\affiliation{Department of Modern Physics, University of Science and Technology of China, Hefei,
Anhui 230026, People's Republic of China}

\begin{abstract}
We present a dynamic approach to study the quantum discord and classical correlation.
By local filtering operation, the evaluation of quantum discord is closely related to quantum channel and channel capacity.
As a consequence, the traditional optimization over horizontal-or-vertical von Neumann measurements is replaced by that over horizontal-and-vertical three-element POVM measurement, from which more rigorous results of quantum discord are obtained.
\end{abstract}


\maketitle

%..........................................................................

Among various measures of quantum correlation, quantum discord (QD) \cite{Ollivier.PhysRevLett.88.017901.2001} has attracted considerable attention.
Many works have been devoted to quantifying and interpreting this measure
\cite{Oppenheim.PhysRevLett.89.180402.2002,
      Vedral.PhysRevLett.90.050401.2003,
      Koashi.PhysRevA.69.022309.2004,
      Horodecki.PhysRevA.71.062307.2005,
      Piani.PhysRevLett.100.090502.2008,
      Luo.PhysRevA.77.022301.2008,
      Datta.PhysRevLett.100.050502.2008,
      Shabani.PhysRevLett.102.100402.2009,
      Ali.PhysRevA.81.042105.2010,
      Modi.PhysRevLett.104.080501.2010,
      Streltsov.PhysRevLett.106.160401.2011,
      Piani.PhysRevLett.106.220403.2011}.
However, the definition of QD involves an optimization procedure over the set of POVM measurements, and thus attacking the general case is a formidable task.
Even for the general 2-qubit states,
the exact expression of QD has not been obtained.
Restricted in the case of 2-qubit X states and von Neumann measurement, Ali presented an analytic method \cite{Ali.PhysRevA.81.042105.2010}:
the direction of the optimal measurement is either vertical (i.e., measuring the Pauli operator $\sigma_z$) or horizontal.
If the X state takes the canonical form, the horizontal measurement is in fact measuring $\sigma_x$.
Recently, it has been pointed out that there exist some irregular states for which the direction of the optimal von Neumann measurement is neither horizontal nor vertical, meaning that this problem has not been solved completely even in the restrict case \cite{Lu.PhysRevA.83.012327.2011,
      *Girolami.PhysRevA.83.052108.2011,
      *Chen.arXiv.1102.0181}.
Meanwhile, there are some works devoted to finding more rigorous results of QD by considering more general POVM measurement \cite{Galve.arXiv.1107.2005}.

In studying QD, the key point is to optimize the Holevo quantity of the postmeasurement ensemble for one subsystem, say system $A$, over all POVM measurements performed on the other subsystem $B$, which gives rise to the classical correlation.
It implies that there is a close relation between QD, in particular classical correlation and the channel capacity.
Roughly speaking, if a channel capacity is achieved with an ensemble containing three or more states \cite{King.PhysRevLett.88.057901.2002}, we can immediately conclude that there must be some states for which the POVM measurement with three or more elements is optimal to acquire the QD or classical correlation.
In the context of 2-qubit state, this situation implies that the von Neumann measurement is not optimal.

We present in this paper a dynamic method to investigate the 2-qubit state QD.
Together with the geometric picture presented in \cite{Shi.NJP.13.073016.2011}, we establish an infinitely-many to one correspondence between the states and the quantum channels.
The characteristic of the channel in turn determines whether or not the optimal measurement will allow for three or four POVM elements.
In this framework, the traditional optimization over horizontal-or-vertical measurements is replaced by that over the horizontal-and-vertical measurements. It turns out that the irregular states can be explained consistently.

Furthermore, we think that the extension of the sudden transition phenomenon presented in \cite{Mazzola.PhysRevLett.104.200401.2010} to general X states should be reconsidered carefully. We have found that for many X states the transition is smooth rather than sudden.
Then we conjecture that, except for a special class of states, there is no sudden transition of QD for general X states undergoing two-sided phase damping channel.

We consider a bipartite quantum state $\rho^{AB}$, and POVM measurement on $B$ with a set of rank one operation elements $M_k\geqslant0$, $\sum_kM_k=\mathbbm1$.
Then QD is defined as $\mathcal{Q}=\mathcal{I}-\mathcal{C}$.
Here $\mathcal{I}$ is quantum mutual information,
$\mathcal{I}=S(\rho^A)+S(\rho^B)-S(\rho^{AB})$, with $\rho^{A(B)}=\Tr_{B(A)}\rho^{AB}$ being the reduced states and $S$ denoting von Neumann entropy.
The classical correlation $\mathcal{C}$ is given by
\begin{equation}
  \mathcal{C}=\max_{\{M_k\}}\Big[S(\rho^A)-\sum_k p_k S(\rho_k^A)\Big],
\end{equation}
where $p_k=\Tr[\rho^{AB}(\mathbbm 1\otimes M_k)]$ and
$\rho^A_k=\Tr_B[\rho^{AB}(\mathbbm 1\otimes M_k)]/p_k$.

The main idea in the method is fixing the steering ellipsoid and moving the reduced state.

\hypertarget{step i}{(i)} Given a 2-qubit state $\rho^{AB}$, we express it as the unnormalized form $(\Lambda\otimes F)\rho^{AB}$, where $\Lambda$ is a quantum channel acted on qubit $A$, and $F$ a filtering operation on $B$.

\hypertarget{step ii}{(ii)} The channel $\Lambda$ determines the quantum steering ellipsoid $\frak{E}$.
We prove that the form of $\frak{E}$ is invariant under any local filtering operation on $B$.

\hypertarget{step iii}{(iii)} Return to X state $\rho^{AB}$ in the canonical form.
Having obtained the channel $\Lambda$ associated with $\rho^{AB}$, we transform Bell state $\Phi=\ket{\Phi}\bra{\Phi}$ with $\ket{\Psi}=\frac{1}{2}(\ket{00}+\ket{11})$ as
$\Phi\to [\Lambda\otimes\Xi(\xi)]\Phi$, with $\Xi(\xi)$ a parameterized local filtering operation.

\hypertarget{step iv}{(iv)} Denote by $\sigma^{AB}(\xi)$ the normalized form of the state obtained in step (iii).
By selecting suitable filtering operator $\Xi(\xi)$, the $\sigma^{AB}(\xi)$ remains in the X form, and furthermore we can translate the position of $\sigma^A(\xi)=\Tr_B\sigma^{AB}(\xi)$, in the geometric picture, along vertical direction between the lower and the upper apex of the ellipsoid $\frak{E}$.

\hypertarget{step v}{(v)} For each $\sigma^A(\xi)$, we calculate two Holevo quantities $\chi_{\leftrightarrow}(\xi)$ and $\chi_{\updownarrow}(\xi)$, which correspond to measuring $\sigma_x$ and $\sigma_z$, respectively, of qubit $B$. Compare the two Holevo quantities.
We conclude that if $\chi_{\leftrightarrow}(\bar{\xi})$ intersects $\chi_{\updownarrow}(\bar{\xi})$ at a specific $\bar{\xi}$, then there are in the neighborhood of $\bar\xi$ the states for which the optimal measurement is three- or four-element of POVM measurement.

Now we begin with step \hyperlink{step i}{(i)}. Usually we express the 2-qubit state $\rho^{AB}$ in the Hilbert-Schmidt space as
$R=2\Upsilon(\rho^{AB})^R\Upsilon^T$ \cite{Verstraete.PhysRevA.64.010101.2001}, where the superscript $R$ denotes the reshuffling transformation: $\ket{ij}\bra{i'j'}\to\ket{ii'}\bra{j'j}$, and the superscript $T$ means transposition.
For the sake of clarity, we consider X states in the canonical form.
The canonical X state $\rho^{AB}$ and the unitary matrix $\Upsilon$ are given by
\begin{equation} \label{X state and Upsilon}
  \rho^{AB}=
  \begin{pmatrix}
    \;a\; & \;0\; & \;0\; & \;u\; \\
    0 & b & v & 0 \\
    0 & v & c & 0 \\
    u & 0 & 0 & d
  \end{pmatrix}, \,
  \Upsilon=\frac{1}{\sqrt{2}}
    \begin{pmatrix}
      1  &  0  &  0  &  1 \\
      0  &  1  &  1  &  0 \\
      0  &  i  &  -i  & 0 \\
      1  &  0  &  0  &  -1
    \end{pmatrix},
\end{equation}
where $a+b+c+d=1$, $u,v\geqslant 0$, $u^2\leqslant ad$ and $v^2\leqslant bc$.

Let filtering operation take the form $F=\diag(\sqrt{a+c},\sqrt{b+d})$.
The corresponding superoperator is $\Lambda_F=F\otimes F^*=F\otimes F$.
To determine the channel $\Lambda$, we apply local operation $\Lambda\otimes\Lambda_F$ on Bell state $\Phi$ and write
$(\rho^{AB})^R=2\Lambda\,\Phi^R\,\Lambda_F^T$,
where the factor $2$ comes from normalization.
By noting $\Phi^R=\mathbbm1/2$, we have $\Lambda=(\rho^{AB})^R\Lambda_F^{-T}$.
Thus step \hyperlink{step i}{(i)} is finished.

It is helpful to express the channel $\Lambda$ in Heisenberg representation:
\begin{equation} \label{LA for X state}
\begin{split}
   & L=\Upsilon\Lambda\Upsilon^\dag \\
   = & \begin{pmatrix}
       1 & 0 & 0 & 0 \\
       0 & \frac{u+v}{\sqrt{(a+c)(b+d)}} & 0 & 0 \\
       0 & 0 & \frac{u-v}{\sqrt{(a+c)(b+d)}} & 0 \\
       \frac{ab-cd}{(a+c)(b+c)} & 0 & 0 & \frac{ad-bc}{(a+c)(b+d)}
       \end{pmatrix}.
\end{split}
\end{equation}

The effect of $L$ on one qubit state is the transformation of Bloch sphere to a ellipsoidal surface, which is given by
\begin{equation} \label{X state Eequation}
  \frac{x^2}{\ell_1^2}+\frac{y^2}{\ell_2^2}+\frac{(z-z_0)^2}{\ell_3^2}=1,
\end{equation}
where
\begin{align*}
  & \ell_1=\frac{u+v}{\sqrt{(a+c)(b+d)}}, \quad
    \ell_2=\frac{|u-v|}{\sqrt{(a+c)(b+d)}}, \\
  & \ell_3=\frac{|ad-bc|}{(a+c)(b+d)}, \quad
    z_0=\frac{ab-cd}{(a+c)(b+d)}.
\end{align*}
It is just the quantum steering ellipsoid $\frak{E}$ which we have introduced to discuss quantum discord.

\begin{figure}[tbph]
\begin{center}
\includegraphics[width=0.5\columnwidth]{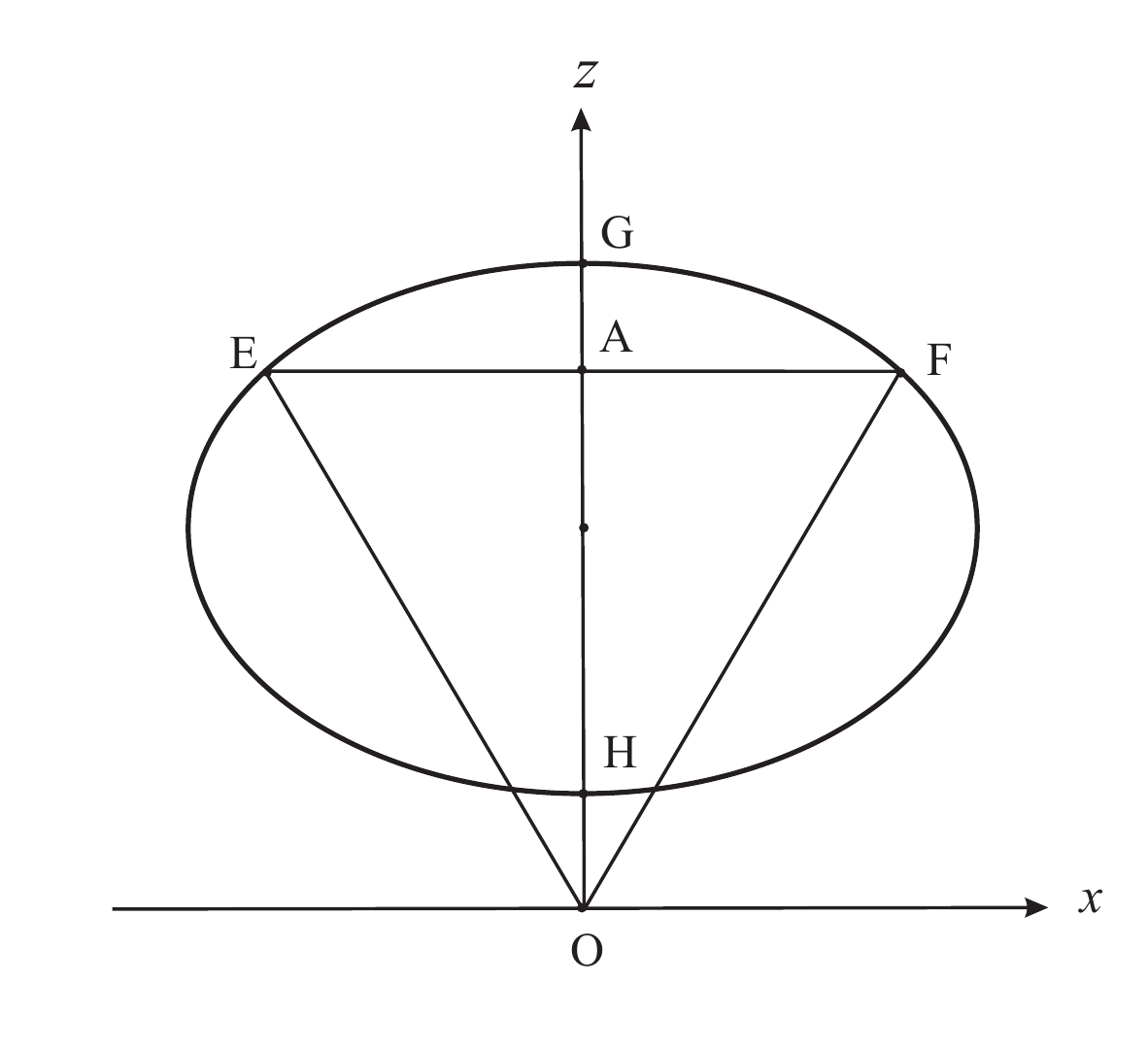}
\end{center}
\caption{Schematic plot of the largest vertical intersection given by $(x/\ell_1)^2+[((z-z_0)/\ell_3]^2=1$. Two pairs of points, $(E,F)$ and $(G,H)$, correspond to the horizontal and vertical decomposition of $\rho^A$, respectively.}
\label{fig:LE for X states}
\end{figure}

In Fig. \ref{fig:LE for X states}, we plot schematically the largest vertical intersection of $\frak{E}$.
The reduced state is represented by the point $A$.
The $\sigma_x$ measurement on qubit $B$ will induce the (horizontal) decomposition of $\rho^A$ as $\mathcal{E}^A_{\leftrightarrow}=\{p_E,\rho^A_E;\,p_F,\rho^A_F\}$ with $p_E=p_F=1/2$ and $S(\rho^A_E)=S(\rho^A_F)$.
The $\sigma_z$ measurement will give rise to the (vertical) ensemble for $\rho^A$:
$\mathcal{E}^A_{\updownarrow}=\{p_G,\rho^A_G;\,p_H,\rho^A_H\}$.
See \cite{Shi.NJP.13.073016.2011} for detail.

Now we prove the following lemma.
\begin{lemma} \label{lemma 1}
  The form of the quantum steering ellipsoid is invariant under the local filtering operation on qubit $B$.
\end{lemma}

The equation of the steering equation is given by
$(1\;x\;y\;z)\,\mathfrak{E}\,(1\;x\;y\;z)^T=0$,
where the symbol $\mathfrak{E}$ is also used to denote the $4\times4$ to  symmetric matrix given by
$\mathfrak{E}=R^{-T}\eta R^{-1}$
with $\eta=\diag(1,-1,-1,-1)$.
With arbitrary filtering operation $\Xi$ applied on qubit $B$, we have
\begin{equation*}
    \mathfrak{E}\rightarrow \mathfrak{E}'
    =R^{-T}L_\Xi^{-1}\eta L_\Xi^{-T}R^{-1}.
\end{equation*}
It is easy to see that $L_\Xi\eta L_\Xi^{T}=|\det(\Xi)|^2\,\eta$. Then it follows that the matrix $\frak{E}'$ is proportional to the matrix $\frak{E}$ and that the equation of steering ellipsoid in invariant.

Lemma \ref{lemma 1} holds for any 2-qubit states.
It means that given a steering ellipsoid $\frak{E}$, there are infinitely many states corresponding to it.
The reduced state of qubit $A$ is represented by a point in the interior of $\frak{E}$.
We can move the point by performing the local filtering on qubit $B$.
Step \hyperlink{step ii}{(ii)} is then realized.

We now proceed to step \hyperlink{step iii}{(iii)}.
Define a parameterized filtering operator
$\Xi(\xi)=\diag(\xi,\,\sqrt{1-\xi^2})$
with $\xi\in(0,1)$.
Apply local operation $\Lambda\otimes\Lambda_\xi$ with $\Lambda_\xi=\Xi(\xi)\otimes\Xi^*(\xi)$ on Bell state $\Phi$.
The computable form is given by, after normalization,
\begin{equation} \label{sigma gamma}
  \Phi^R\to
  \big[\sigma^{AB}(\xi)\big]^R=2\Lambda\,\Phi^R\,\Lambda_\xi^T
  =\Lambda\,\Lambda_\xi^T
\end{equation}
The equivalent expression is
\begin{equation} \label{R gamma}
  R_\Phi\to R_{\xi}=2L R_\Phi L_\xi^T,
\end{equation}
with $R_\Phi=2\Upsilon\Phi^R\Upsilon^T$, $L$ given by \eqref{LA for X state} and $L_\xi=\Upsilon\Lambda_\xi\Upsilon^\dag$.
When $\xi=\sqrt{a+c}$, $R_\xi$ becomes the $R$ matrix of the X state \eqref{X state and Upsilon}.
When $\xi$ changing from $0$ to $+1$, the $z$-component of the Bloch vector of $\rho^{A}(\xi)$ changes from $\frac{b-d}{b+d}$ to $\frac{a-c}{a+c}$, which are the two apexes in vertical direction.
So step \hyperlink{step iii}{(iii)} is realized.

Step \hyperlink{step iv}{(iv)} is straightforward.
The steering ellipsoid $\frak{E}$ is given by \eqref{X state Eequation}.
Any point on $z$ axis between point $G$ and point $H$, which represents the reduced state $\sigma^A(\xi)$, is in one-to-one correspondence to $\xi\in(0,1)$.
The two Holevo quantities are given by
\begin{align}
  \chi_\leftrightarrow(\xi)
   & =S[\sigma^A(\xi)]-p_ES[\sigma^A_E(\xi)]-p_FS[\sigma^A_F(\xi)] \nonumber\\
   & =S[\sigma^A(\xi)]-S[\sigma^A_E(\xi)], \\
  \chi_\updownarrow(\xi)
   & =S[\sigma^A(\xi)]
      -p_GS[\sigma^A_G(\xi)]-p_HS[\sigma^A_H(\xi)]
\end{align}
Note that $\chi_\leftrightarrow(0)=\chi_\updownarrow(0)$ and
$\chi_\leftrightarrow(1)=\chi_\updownarrow(1)$.
It can be verified that there is at most one intersection point of $\chi_\leftrightarrow(\xi)$ and $\chi_\updownarrow(\xi)$ except the trivial $\xi=0,1$.

\begin{figure}[bpth]
\begin{center}
  \includegraphics[width=0.4\textwidth]{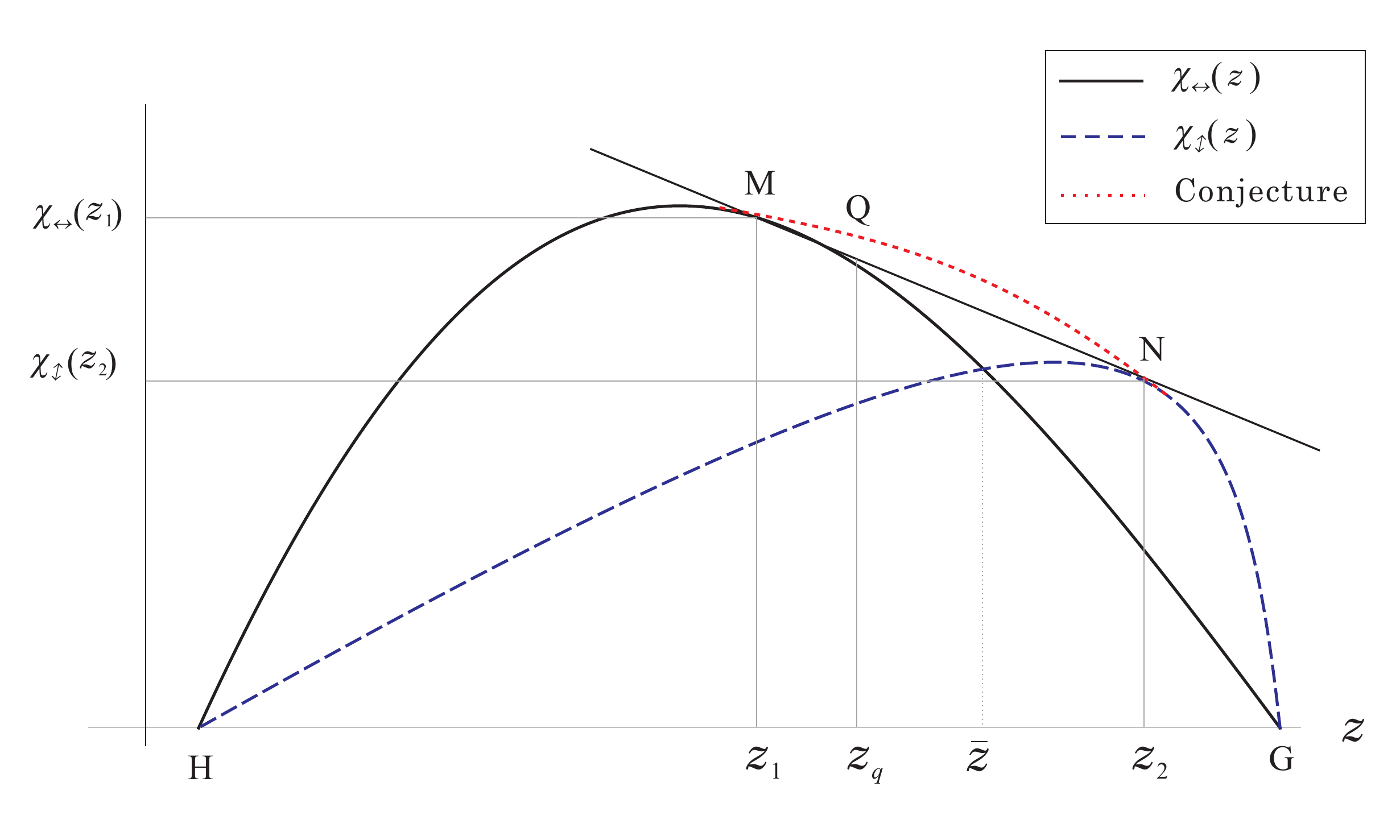}
  \caption{Schematic plot of $\chi_\leftrightarrow(z)$ and $\chi_\updownarrow(z)$.
  For any $z\in(z_1,z_2)$, there exits a three- or four-element ensemble for $\sigma^A(z)$ which gives rise to the more optimal Holevo quantity.}\label{fig: CC and LF}
\end{center}
\end{figure}

Now it is time to verify the conclusion presented in step \hyperlink{step v}{(v)} to complete our argument.
We have to compare two functions $\chi_{\leftrightarrow}(\xi)$ and $\chi_{\updownarrow}(\xi)$.
For the sake of clarity, we replace the argument of the Holevo quantities by the position of $\sigma^A(\xi)$ in the geometric picture, denoted by $z$.

We plot schematically in Fig. \ref{fig: CC and LF} the functions $\chi_{\leftrightarrow}(z)$ and $\chi_{\updownarrow}(z)$.
Assume that they intersect at $\bar{z}$.
We plot a line tangent to both curves with tangent points denoted by $M$ and $N$.
Corresponding to $M$ and $N$, the reduced states are $\sigma^{A}(z_1)$ and $\sigma^{A}(z_2)$, respectively.
It can be seen that $\bar{z}\in(z_1,z_2)$.
For $\sigma^A(z_1)$, the horizontal decomposition,
$\mathcal{E}_{\leftrightarrow}^A(z_1)
=\{p_E,\sigma^A_E(z_1);\;p_F,\sigma^A_F(z_1)\}$,
gives rise to $\chi_{\leftrightarrow}(z_1)$.
Here, both probabilities $p_E$ and $p_F$ are in fact equal to $1/2$.
For $\sigma^A(z_2)$, the vertical decomposition,
$\mathcal{E}_{\updownarrow}^A(z_2)
=\{p_G,\sigma^A_G(z_2);\;p_H,\sigma^A_H(z_2)\}$,
gives rise to $\chi_{\updownarrow}(z_2)$.

Now consider a state $\sigma^A(z_q)=q\sigma^A(z_1)+(1-q)\sigma^A(z_2)$ with $q\in(0,1)$.
We can see that $z_q=qz_1+(1-q)z_2$.
Let us consider the 4-state ensemble for $\sigma^A(z_q)$:
\begin{equation}
\begin{split}
  \mathcal{E}^A_4(z_q)=&\Big\{q\,p_E,\sigma_E^A(z_1);\;
                        q\,p_F,\sigma_E^A(z_1); \\
  &\; (1-q)p_G,\sigma_G^A(z_2);\;(1-q)p_H,\sigma_H^A(z_2)\Big\}.
\end{split}
\end{equation}
We calculate the Holevo quantity of the ensemble $\mathcal{E}_4^A(z_q)$.
\begin{equation} \label{chi4}
\begin{split}
  \chi_4(z_q)=q\chi_{\leftrightarrow}(z_1)
    +(1-q)\chi_{\updownarrow}(z_2)+S[\sigma^A(z_q)] & \\
      -qS[\sigma^A(z_1)]-(1-q)S[\sigma^A(z_2)] & .
\end{split}
\end{equation}
The linear combination of the last three terms at the right-hand side of \eqref{chi4} is strictly larger than zero due to the concavity of entropy.
As for the first two terms,
it is easy to see that for any $q\in(z_1,z_2)$
the sum $q\chi_{\leftrightarrow}(z_1)+(1-q)\chi_{\updownarrow}(z_2)$ is strictly larger than both
$\chi_{\leftrightarrow}(q)$ and $\chi_{\updownarrow}(q)$.
It turns out that $\chi_4(z_q)>\chi_{\leftrightarrow}(q)$ and
$\chi_4(z_q)>\chi_{\updownarrow}(q)$ for $q\in(z_1,z_2)$.
Thus we prove that, for states in this interval, four-element POVM measurement is optimal.
In fact, four-element ensemble $\mathcal{E}^A_4$ can be simplified to three-element ensemble $\mathcal{E}^A_3$, in which one element corresponds to one of two the vertical apexes of $\frak{E}$, and the other two elements correspond to two intersection points of a vertical line and $\frak{E}$.
From this perspective,
we conjecture that the Holevo quantity $\chi(z)$ could be a smooth curve.

In the context of evaluating QD or classical correlation of X states,
the optimal three-element POVM measurement is such that one measurement operator is $\ket{0}\bra{0}$ or $\ket{1}\bra{1}$ and the other two are $\ket{+x}\bra{+x}$ and $\ket{-x}\bra{-x}$ respectively.
Applying this type of measurement to the irregular state presented in \cite{Lu.PhysRevA.83.012327.2011},we see a more rigorous result of QD.

The dynamic method presented above is concerned with the situation that the quantum steering ellipsoid is fixed while the reduced state of one qubit is pushed up and down by the filtering operation applied on the other qubit.
On the contrary, we can fix the reduced state and deform the ellipsoid.
This process can be realized by using phase damping channel.

When two-sided phase damping channel is acted on the state $\rho^{AB}$, the time evolution is given by
\begin{equation*}
    \rho^{AB}(t)=\sum_{i,j=1}^{2}(K_i\otimes K_j)\rho(K_i\otimes K_j)^\dag,
\end{equation*}
where $K_1=\mathrm{diag}(1,\gamma)$ and $K_2=\mathrm{diag}(0,\sqrt{1-\gamma^2})$ are Kraus operators representing phase damping channel, and $\gamma=e^{-\Gamma t}$ with $\Gamma$ the phase damping rate.
Here we assume that qubits $A$ and $B$ endure the same noisy environment. At initial time $t=0$ the steering ellipsoid $\frak{E}(0)$ is given by \eqref{X state Eequation}. At time $t>0$, the ellipsoid is transformed to $\frak{E}(t)$, which is expressed by
\begin{equation*}
    \frac{x^2}{(\gamma^2\ell_1)^2}+\frac{y^2}{(\gamma^2\ell_2)^2}
    +\frac{(z-z_0)^2}{\ell_3^2}=1.
\end{equation*}
That is, with $\gamma$ decreasing from $1$ to $0$, the radius of the ellipsoid along $x$ axis and that along $y$ axis decrease continuously from $l_1$ and $l_2$ respectively to zero, whereas the radius along $z$ axis remains the same.
The reduced state $\rho^A$ is not affected by phase damping.
For the state $\rho^{AB}(t)$ at time $t$, we have two Holevo quantities: $\chi_{\leftrightarrow}(t)$ and $\chi_{\updownarrow}$.
Note that $\chi_{\updownarrow}$ is independent of time $t$.
If initially $\chi_{\leftrightarrow}(0)>\chi_{\updownarrow}$, there must be a critical time $\bar{t}$ on which $\chi_{\leftrightarrow}(\bar{t})=\chi_{\updownarrow}$.
It follows that $\chi_{\leftrightarrow}(t)<\chi_{\updownarrow}$ for all $t>\bar{t}$.
Horizontal-or-vertical optimization tells us that at time $\bar{t}$ there is a sudden transition from classical to quantum decoherence regime. However, the horizontal-and-vertical optimization over 3-element POVM measurements shows that this sudden transition may not exist.

\begin{figure}[bpth]
\begin{center}
  \includegraphics[width=0.4\textwidth]{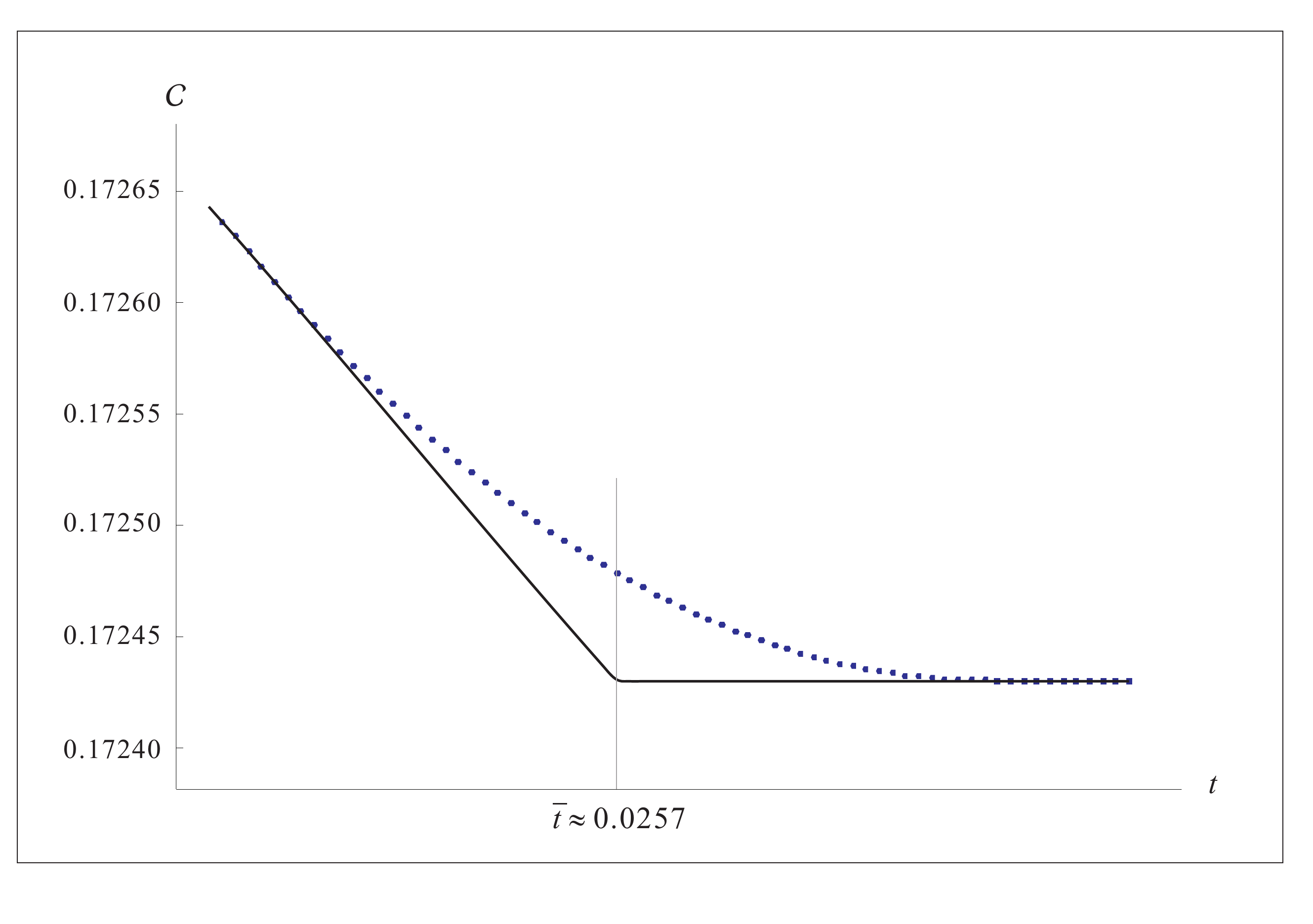}
  \caption{Time evolution of the classical correlation of the 2-qubit state, with the element of $\rho^{AB}$ given by $a=0.4875$, $b=0.1625$, $c=0.0875$, $d=0.2625$, $u=0.3354$ and $v=0.1118$, undergoing two-sided phase damping channel. Solid line corresponds to the result coming from horizontal-or-vertical von Neumann measurement, while the dotted line to the result obtained by three-element POVM measurement.}\label{fig:PD}
\end{center}
\end{figure}

For example, let us consider the state presented in \cite{Li.PhysRevA.83.022321.2011}.
We compare in Fig. \ref{fig:PD} two numerical results of the classical of the state undergoing two-sided phase damping channel with $\Gamma=0.01$. It is shown that the three-element POVM measurement is more optimal than the traditional von Neumann measurement.

A special case must be pointed out.
If the center of the ellipsoid $\frak{E}$ is on the origin point, the sudden transition must occur as long as $\chi_{\leftrightarrow}(0)>\chi_{\updownarrow}$.
In this case, when $\chi_{\leftrightarrow}(\bar{t})$ is equal to $\chi_{\updownarrow}$ at the critical time $\bar{t}$, the largest vertical intersection of the $\frak{E}(\bar{t})$ is a circle. It turns out that $\chi_{\leftrightarrow}(\bar{t},z)$ is identical  $\chi_{\updownarrow}(z)$ for all $z$ between the lower and the upper apexes of the circle, and there is no region in which the four- or three-state ensemble can yields a more optimal Holevo quantity.

In conclusion, we present a dynamic approach to study quantum correlation and classical correlation.
With the the dynamic process illustrated explicitly in the geometric picture, we find that the QD can be studied from the viewpoint of quantum channel and channel capacity.
As a consequence, the traditional optimization over von Neumann measurements is shown to be not sufficient to give rigorous results.

This work was supported by National Nature Science Foundation of China, the CAS, and the National
Fundamental Research Program 2007CB925200.

\bibliography{QDandLF}

%-----------------------------------------------------------------------------
\clearpage
\end{document}